\documentclass{optica-article}
\journal{opticajournal}
\articletype{Research Article}
\usepackage{float}
\usepackage{subcaption}

\usepackage{lineno}

\usepackage[linesnumbered,ruled,vlined]{algorithm2e}
\makeatletter
\providecommand{\doi}[1]{%
  \begingroup
    \let\bibinfo\@secondoftwo
    \urlstyle{rm}%
    \href{http://dx.doi.org/#1}{%
      doi:\discretionary{}{}{}%
      \nolinkurl{#1}%
    }%
  \endgroup
}
\makeatother
\usepackage[nameinlink,noabbrev,capitalize]{cleveref}

\renewcommand{\cref}{\Cref}

\let\oldexp\exp
\renewcommand{\exp}[1]{\oldexp{\left( {#1} \right)}}

\begin{document}

	\title{Lensless and Lossless HoloVAM}
	
	\author{\authormark{*}Andreas Erik Gejl Madsen and Jesper Gl\"uckstad}

	\address{SDU Centre for Photonics Engineering\\University of Southern Denmark\\DK-5230 Odense M, Denmark}
	
	\email{\authormark{*}gejl@mci.sdu.dk}
	
\begin{abstract*}
We report the first successful fabrication of three-dimensional models using our fully lensless holographic volumetric additive manufacturing (HoloVAM) platform. In this configuration, tomographic light fields are generated directly from a phase-only spatial light modulator (SLM) and delivered into a rotating vial of photopolymer without any imaging optics, relays, or index-matching bath. Building on the HoloTile framework for tiled Fourier holography and point-spread function (PSF) shaping, the system creates volumetric dose distributions with high photon efficiency and well-controlled axial propagation. Using a simple acrylate resin formulation and a minimalized optical train, we demonstrate reproducible fabrication of complex geometries. These results establish lensless HoloVAM as a practical and mechanically minimal route to volumetric fabrication, opening a new pathway toward compact and application-flexible VAM devices.
\end{abstract*}

\section{Introduction}
\noindent
Volumetric additive manufacturing (VAM) \cite{kelly_computed_2017,loterie_high-resolution_2020,madridwolff_controlling_2022} enables rapid, support-free fabrication by integrating optical dose throughout a rotating photocurable volume. Most existing systems rely on amplitude-modulated projection and multi-element refractive optical trains, large étendue light sources, significant optical power budgets, and mechanical complexity that limits compactness and robustness.
Recent demonstrations of holographic T-VAM \cite{alvarez-castano_holographic_2025,madsen_digital_2024,gluckstad_holotile_2024} have shown that phase-based light engines can suppress speckle, extend axial propagation, and improve light-transport efficiency through the tiled Fourier holography and PSF-shaping strategies of the HoloTile framework \cite{gluckstad_holographic_2022,gluckstad_holographic_2023,madsen_holotile_2022,gluckstad_comparing_2023,gluckstad_holotile_2024-1,madsen_axial_2025,madsen_generalized_2024,madsen_holotile_2025}.

\begin{figure}[!htb]
	\centering
	\includegraphics[width=0.8\textwidth]{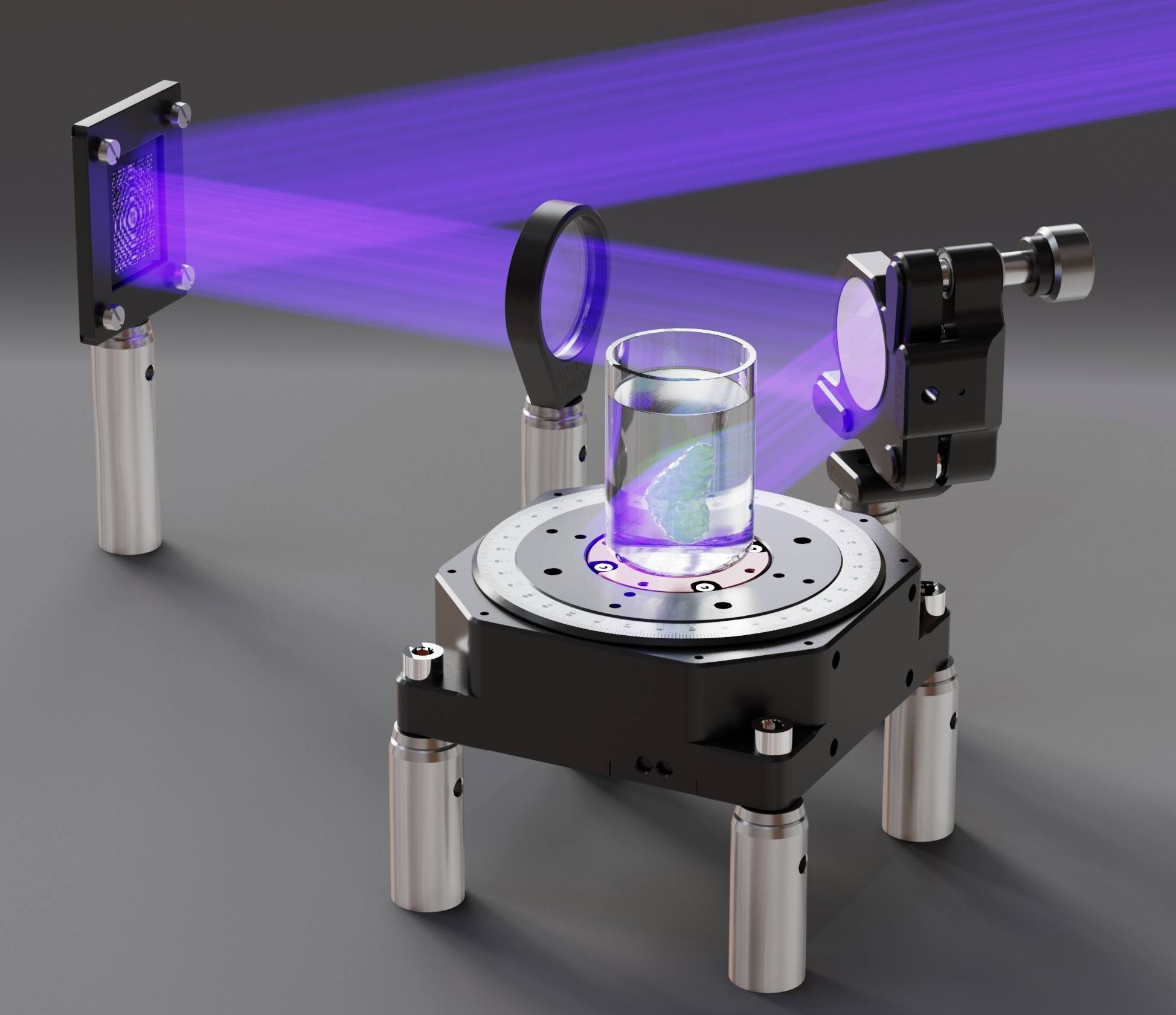}
  \captionsetup{width=\linewidth}
	\caption{Conceptual illustration of the lensless HoloVAM principle. A phase-only SLM is illuminated by an expanded and collimated laser beam. The SLM modulates the wavefront, allowing the necessary projection patterns to form due to the free-space propagation between the SLM and the vial in which a model is cured. A linear polarizer is inserted following the SLM to ensure the passage of only SLM-modulated light. Illustration credit: Sammy Florczak \cite{bernal_road_2025,zandrini_breaking_2023}.}
	\label{fig:concept}
\end{figure}

In this brief report, we show that the same holographic route can be reduced to a minimal functional form: full tomographic curing with no refractive optics in the projection path \cite{madsen_holovam_2026}. In the lensless HoloVAM configuration presented here, a collimated single-mode laser, a phase-only SLM, and a digitally applied quadratic lens phase generate the volumetric exposure that polymerizes the target geometry. Despite the absence of relay lenses or index-matching baths, the system fabricates centimeter-scale, topologically complex solids in under 30 seconds.
These results indicate that full tomographic fabrication can be achieved through holography alone, as illustrated in \cref{fig:concept}. In doing so, lensless HoloVAM directly addresses long-standing barriers in volumetric manufacturing. In particular, the dependence on large, high-power, safety-critical laser projection systems, opening a realistic path to compact, rapid, precise fabrication for biofabrication workflows, organoid and tissue manufacture, patient-specific components in audiology and dental care, and even fabrication in microgravity environments.

\section{Holographic Volumetric Additive Manufacturing}
\noindent
The field of volumetric manufacture has repeatedly emphasized the appeal of a truly volumetric light field capable of solidifying an entire object at once. Conceptually, this would require describing a three-dimensional optical distribution containing far more information than what any two-dimensional modulator can encode. As earlier analyses has noted \cite{somers_holographic_2024,somers_physics_2023}, an ultrafast, fully volumetric exposure remains an elegant but seemingly physically unattainable limit.
\newpage
\noindent
Tomographic approaches emerged as a practical solution to this mismatch. By decomposing the target volume into many angular views, they transform the 3D information into a sequence of feasible 2D patterns, allowing the object to cure through cumulative dose. Yet as long as these patterns are produced by amplitude imaging, they inherit the limits of imaging systems: étendue-limited depth-of-focus, inability to control the axial structure of the light, and low photon efficiency due to sparse projection patterns. Consequently, to force enough cumulative light into the fabrication volume to cure the target volume in any reasonable time, large multimode light sources have become the de-facto standard.

Holography breaks this dependency. Because phase modulation and Fourier holography allows every pixel to contribute to every point in the projected field \cite{madsen_comparison_2022}, the projection is no longer simply a relayed image of the display, but rather a wavefront that can be engineered with specific capabilities. 
The tomographic framework remains as a method of breaking down the 3D volume down to a series of 2D projections, but the light engine is fundamentally transformed. Phase encoding introduces the freedom to design the axial response of the beam, extending its collimated propagation length through Bessel-like or self-healing modes, or compensating for vial-induced aberrations. It also collapses the power requirements by more than an order of magnitude, enabling the use of operator-safe, low-power single-mode sources.

In scattering or turbid media, the wavefront modulating nature of holography is equally important. Amplitude-modulated implementations primarily rely on ballistic photons, and can compensate their attenuation only up to modest cell densities and beyond this regime, the amount of available ballistic photons reduces drastically and the projected patterns lose fidelity. 
Holography provides a broader toolkit that may counteract this. At low turbidity, the wavefront can be pre-aberrated so that the ballistic component arrives undistorted. 
At higher, more physiologically relevant cell densities, the holographic projection can be decomposed into scattering-invariant modes (SIMs) that provide partial robustness even when propagating through scattering media. In this way, phase-encoded HoloVAM can exploit both the surviving ballistic light and a useful subset of scattered light, maintaining dose integrity where amplitude-based approaches fall short.

In this view, holographic VAM represents a promising evolution beyond imaging-based approaches. The lensless implementation demonstrated here represents a minimal expression of this architecture, consisting only of a laser, polarizer, and phase modulator, yet capable of complex volumetric fabrication.

\section{Results}
\noindent
The lensless HoloVAM configuration produces consistent three-dimensional dose distributions inside the rotating resin vial. The axially extended Bessel-like beams, encoded through the HoloTile phase architecture, maintains their beam profile throughout the full vial diameter, consistent with their  non-diffracting behaviour and validated experimentally using fluorescent tracer suspensions (fluorescent powder in water).

Reproducible curing is achieved with total exposure durations of 25-30 seconds, corresponding to approximately two full mechanical rotations of the vial. During this exposure cycle, cumulative dose is deposited in the target volume despite modest background exposure originating from SLM dead-space and residual background speckle noise. The PSF-shaped HoloTile holograms reduce this background noise substantially compared to conventional computer generated holography (CGH), but a small dose bias remains present. Work is ongoing to tune temporal multiplexing and sinogram optimization to further compensate for this background dose.

\begin{figure}[t]
	\centering
	\includegraphics[width=\textwidth]{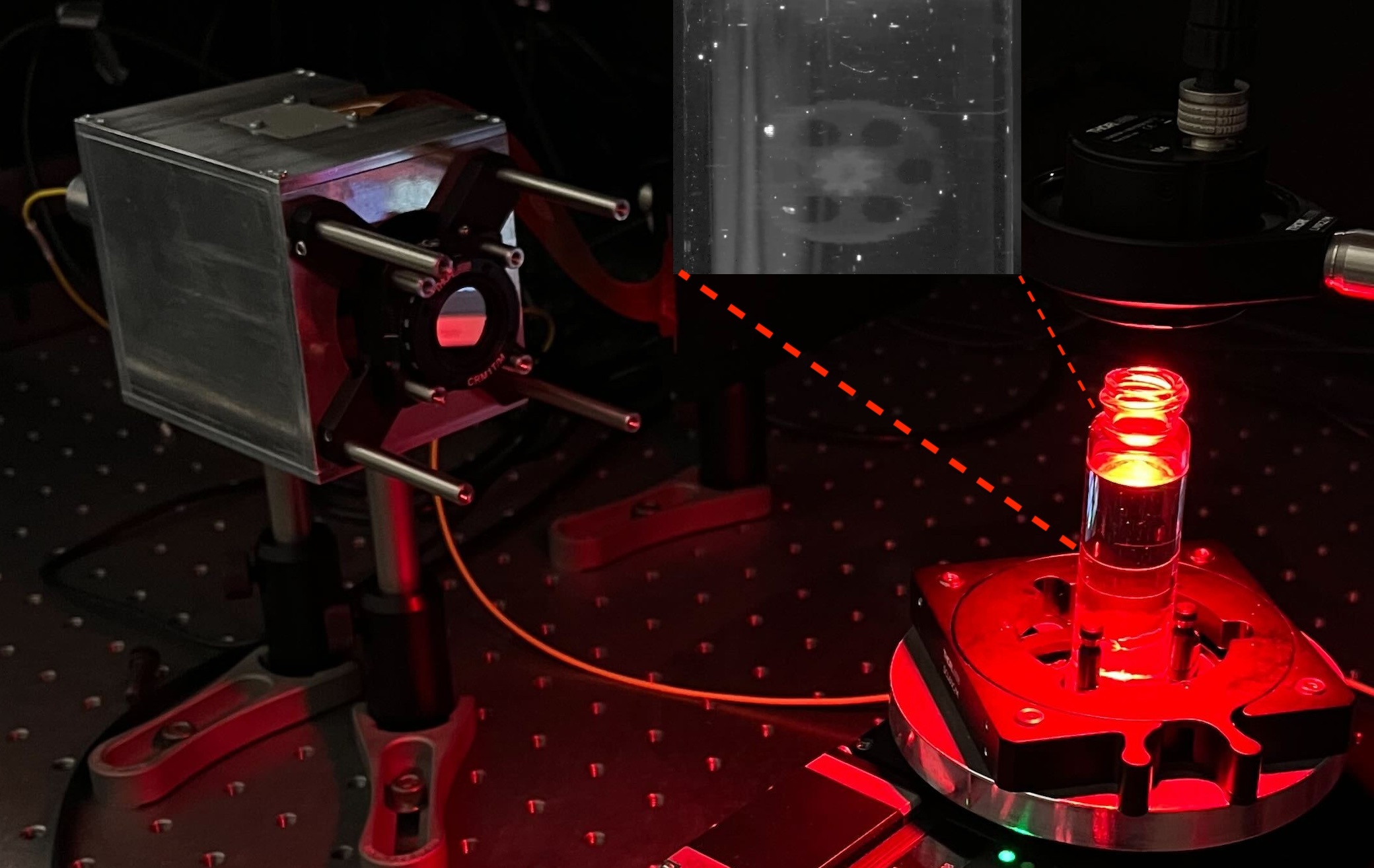}
  \captionsetup{width=\linewidth}
	\caption{Lensless HoloVAM projection setup used for volumetric fabrication experiments. A single-mode 405 nm fiber-coupled diode laser is expanded onto a phase-only spatial light modulator, housed in a compact enclosure. The SLM encodes HoloTile holograms that generate axially extended Bessel-like beams, which propagate directly into a rotating cylindrical resin vial. A monitoring camera positioned off-axis to the vial captures the curing process (grayscale inset).}
	\label{fig:composite}
\end{figure}

Using the final pipeline as described in \cref{sec:methods}, several geometries were fabricated at centimetre scale (maximum dimension 12 mm), an example of which is seen in the inset in \cref{fig:composite} as captured in-vial by the monitoring camera, being a model of a gear sprocket.
All three models were reproduced with high geometric fidelity and minimal large-scale distortion. The gear exhibited resolved teeth, indicating that the system supports well-defined positive and negative features at relevant scales.

As is also common in conventional VAM, the fabricated models exhibited periodic surface modulations, also known as striations \cite{alvarez-castano_holographic_2025,rackson_latent_2022-1}. The spatial frequency of the striations align with the characteristic output pixel spacing.
Importantly, this artefact is not unique to lensless HoloVAM and several mitigation strategies (PSF engineering, temporal multiplexing, flood curing, etc.) are discussed in \cref{sec:discussion}.

Fabrications of all three models were repeated with consistent qualitative outcomes in shape fidelity, striation pattern, and global dimensional accuracy.
The overall performance (feature resolution, volumetric uniformity, and reproducibility) demonstrates that a lensless holographic engine is sufficient for practical volumetric fabrication.

Without the vial-curvature compensation and sinogram optimization described in \cref{sec:computational}, the fabricated models consistently narrowed relative to their digital models. This deformation originates from the inward refraction of the collimated reconstruction beams due to the lack of an index-matching bath. The compensation scheme and iterative optimization pipeline restores the intended lateral dimensions.

\section{Methods}
\label{sec:methods}
\noindent
All exposures were generated using our lensless holographic light engine, based on a single phase-only SLM. A digitally applied quadratic phase establishes the Fourier relationship between the SLM and the resin reconstruction volume without any physical optics. The SLM is therefore the sole wavefront-shaping element, and the 3-dimensional exposure field forms directly inside the rotating vial.

\begin{figure}[!htb]
	\centering
	\includegraphics[width=.9\textwidth]{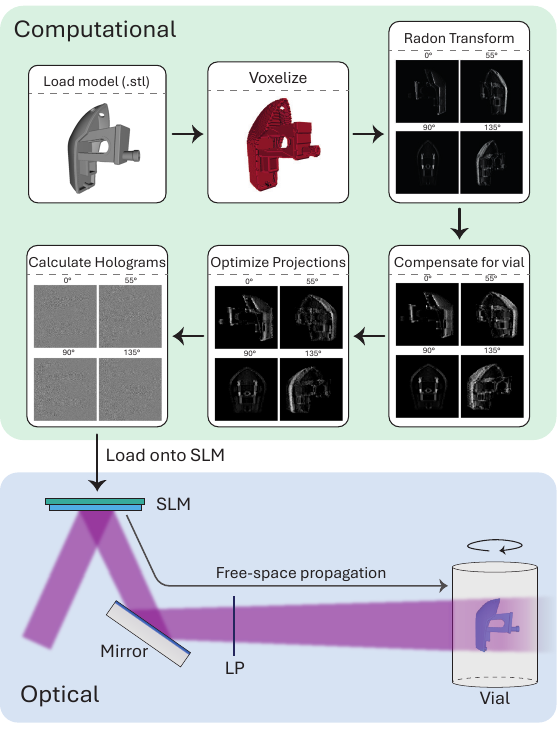}
  \captionsetup{width=\linewidth}
	\caption{Computational and optical workflow of the lensless HoloVAM system. Target meshes are voxelized and converted to angular projections, which are pre-warped for cylindrical refraction and refined through an iterative dose-based optimization. Each corrected projection is converted to a phase-only HoloTile hologram containing both tiling, PSF shaping, and the digitally applied lens phase. The holograms are displayed on the SLM in sync with vial rotation, and, after polarization filtering, propagate freely to form the three-dimensional exposure field directly inside the resin-filled vial.}
	\label{fig:overview}
\end{figure}

In this section, both the computational and optical pipelines are briefly described, going through the preparation for fabrication and the optical system. The whole process is illustrated in \cref{fig:overview}.

\subsection{Computational Pipeline}
\label{sec:computational}
Target geometries (.STL 3D meshes) are voxelized at an isotropic resolution suitable for the intended fabrication size and the sub-hologram resolution available on the SLM. A discrete Radon transform generates the initial angular projections. These projections are first pre-warped to counteract the geometric distortions introduced by cylindrical refraction of the vial shape, the result of which are further refined through an iterative optimization scheme \cite{rackson_object-space_2021,wechsler_wave_2024,nicolet_inverse_2024}.
Each iteration uses a simplified forward dose simulation where rays, corresponding to output pixel locations, traverse through a refractive model of the vial and resin. The simulation accounts for refraction at both air-vial and vial-resin interfaces, Beer–Lambert absorption, and the inherent divergence of rays originating from the Gaussian-illuminated SLM. To refine the projections for optimal curing dynamics, the loss function includes terms to separate the distributions of accumulated dose within and outside the target geometry.

For each optimized projection, corresponding to a specific angular rotation of the vial, a phase-only HoloTile hologram is synthesized, coded with the extended Bessel-like output pixel shape, as well as the digitally applied lens phase. During fabrication, the hologram sequence is displayed on the SLM, synchronized with the vial rotation. At each angular position, the phase pattern generates a full three-dimensional intensity distribution of axially extended Bessel-like beams inside the resin.

\subsection{Optical Pipeline}
\label{sec:optical}
The optical setup consists of a 30 mW, 405 nm pigtail fiber-coupled laser diode whose beam is expanded in a fixed $7.5\times$ beam expander, a phase-only SLM (Holoeye Photonics PLUTO-2.1-UV), a linear polarizer to extract only the phase-modulated light, and the curing vial itself placed on a rotary stage (Zaber Motion X-RSW60C-E03), as seen in \cref{fig:composite}.

Following modulation by the SLM and filtering by the linear polarizer, the light propagates through free space a distance equal to the focal length of the digitally applied lens phase directly into the resin-filled cylindrical vial \cite{gluckstad_new_2023,gluckstad_gabor-type_2024}. No lenses, relay optics, or index-matching components are placed between the SLM and the curing volume.
A separate monitoring channel operates independently of the curing beam. A 630 nm LED array illuminates the vial from above, and an off-axis camera records the curing progression throughout the exposure. 

\subsection{Resin formulation}
All experiments use a photocurable acrylate resin based on dipentaerythritol penta/hexa-acrylate with diphenyl(2,4,6-trimethylbenzoyl)phosphine oxide (TPO) as the photoinitiator. The base resin is first degassed in a heated ultrasonic bath, after which TPO is added to a concentration of approximately 3mM and mixed thoroughly. The homogeneous mixture is then dispensed into cylindrical glass vials and allowed to cool.


\section{Discussion}
\label{sec:discussion}
\noindent
This first demonstration of lensless HoloVAM shows that phase encoding alone is sufficient to generate stable tomographic doses and reproducibly fabricate complex 3D structures within tens of seconds. While this report shows the first demonstrated prototype of the lensless HoloVAM system, the results suggests clear opportunities for performance improvements.

Several improvements follow directly from available SLM hardware. Higher pixel counts will allow for larger sub-holograms and finer spatial features, while reduced pixel dead-space will diminish background illumination and improve dose contrast. 
Striations, which arise from the narrow lateral extent of the Bessel-like beams, can be mitigated through temporal PSF multiplexing, latent-image VAM strategies that conclude with a flood-cure step \cite{rackson_latent_2022-1}, or alternative beam shapes that distribute dose more uniformly, several of which will be explored in subsequent work.
Future optimization schemes can also incorporate the full holographic forward model, enabling compensation for background, speckle, and axial propagation effects rather than correcting only in sinogram space \cite{wechsler_wave_2024, nicolet_inverse_2024}.
Because the system is fully phase programmable, corrections that are currently applied to the projections can, in future, instead be expressed as wavefront pre-aberrations. This approach will allow for phase-compensation of vial curvature, refractive-index inhomogeneity, and scattering media, and could provide a path toward closed-loop correction during curing.

\section{Conclusion}
\noindent
We have shown that tomographic volumetric fabrication can be achieved with our lensless, phase-encoded light engine, HoloVAM. By exploiting holographic projection and engineered PSFs, the system generates extended-propagation, low-divergence beams and deposits dose patterns that support centimetre-scale 3D fabrication in under 30 seconds. These results indicate that lensless HoloVAM offers a practical and compact alternative to amplitude-based imaging approaches, particularly when power efficiency, form factor, or alignment sensitivity are limiting factors.
Although the present demonstration prioritizes simplicity, the architecture provides clear routes for refinement through higher-resolution modulators, improved PSF design, and wave-optical optimization. Overall, this work establishes lensless HoloVAM as a feasible and mechanically minimal platform for future volumetric fabrication technologies.

\section*{Acknowledgements}
\noindent
This work has been supported by Spin-outs Denmark (Translational Postdoc Program) and the Novo Nordisk Foundation (Grand Challenge Program; NNF16OC0021948).

\bibliography{references.bib}

\end{document}